\documentclass[aps,12pt,preprint,superscriptaddress,showpacs]{revtex4}
\usepackage{amssymb,latexsym,amscd,amsmath,epsfig}
\usepackage{longtable}

\def\be{\begin{equation}}
\def\ee{\end{equation}}
\def\ba{\begin{array}}
\def\ea{\end{array}}
\def\bea{\begin{eqnarray}}
\def\eea{\end{eqnarray}}
\def\bi{\begin{itemize}}
\def\ei{\end{itemize}}

\def\half{{\textstyle{1\over2}}}

\begin{document}
\title{Constraining the gravitational binding energy of PSR J0737-3039B using terrestrial nuclear data}
\author{W.~G.~Newton}
\author{Bao-An~Li}
\affiliation{Department of Physics, Texas A\&M University-Commerce, Commerce, Texas 75429-3011, USA}
\date{\today}
\begin{abstract}
We show that the gravitational binding energy of a neutron star of a given mass is correlated with the slope of the nuclear symmetry energy at 1-2 times nuclear saturation density for equations of state without significant softening (i.e., those that predict maximum masses $M_{\rm max} > 1.44M_{\odot}$ in line with the largest accurately measured neutron star mass). Applying recent laboratory constraints on the slope of the symmetry energy to this correlation we extract a constraint on the baryon mass of the lower mass member of the double pulsar binary system, PSR J0737-3039B. We compare with independent constraints derived from modeling the progenitor star of J0737-3039B up to and through its collapse under the assumption that it formed in an electron capture supernova. The two sets of constraints are consistent only if $L \lesssim$ 70 MeV.
\end{abstract}

\maketitle

\section{\label{sec1}Introduction}

The double pulsar system PSR J0737-3039A/B \cite{Burgay2003,Lyne2004} consists of two radio pulsars with masses and spin periods 1.3381$\pm$0.0007$M_{\odot}$, 22.7ms (pulsar A) and 1.2489$\pm$0.0007$M_{\odot}$, 2.77s (pulsar B) in a compact 2.4h orbit with a relatively low eccentricity of $e =$ 0.088. Precision observations of the system are providing excellent tests of a wide range of fundamental physics, including stringent tests of gravity in the strong field regime \cite{Kramer2008}. J0737-3039A/B also provides insight into the evolutionary paths of binary systems that produce double neutron star systems (DNSs).

DNSs originate in binaries where, necessarily, both stars are initially massive enough to undergo supernova (SN) explosions. We shall refer to the supernovae that created pulsars A and B as SNA and SNB respectively. It has been shown \cite{Dewi2002,Ivanova2003,Podsi2005} that the possible evolutionary paths of the binary lead to the immediate precursor of the DNS being a helium star (HeS) - neutron star (NS) system.

The low mass of pulsar B and low eccentricity of the orbit ($e =$ 0.088; estimated initial eccentricity $e_0 \approx$ 0.11-0.12 \cite{Burgay2003}) led to the suggestion that it was formed, not by the collapse of an iron core, but by the collapse of an ONeMg core initiated by electron captures ($e$-captures) onto Mg; that is, an $e$-capture supernova \cite{Nomoto1984}. The gravitational mass of pulsar B is $M_{\rm G} \approx$ 1.25$M_{\odot}$, and its gravitational binding energy may be roughly estimated to be $\approx 0.084(M_{\rm G}/M_{\odot})^2 M_{\odot} =  0.13M_{\odot}$ \cite{Lattimer1989} leading to an estimate of the baryon mass of the neutron star of $M_{\rm B} \approx 1.38M_{\odot}$. This is close to the estimated mass at which an ONeMg core becomes unstable to collapse \cite{Nomoto1984,Podsi2005}. Such a supernova would be relatively symmetric as the timescale for the explosion is expected to be significantly shorter than the timescale for large instabilities to develop \cite{Scheck2004,Podsi2004,Kitaura2006}; this would explain the low eccentricity of the system. Assuming a symmetric SN, the eccentricity implies that the mass expelled from the HeS progenitor during SNB was $\Delta M \approx e_0 (M_{\rm A} + M_{\rm B})$ = 0.32$M_{\odot}$ and the corresponding total mass for that HeS was $\approx 1.56 M_{\odot}$. Such a low mass helium star would form in a HeS-NS binary in which the initial HeS mass was $\approx 3 M_{\odot}$ \cite{Dewi2002,Ivanova2003}, placing it in the mass range of helium stars expected to end their life in an $e$-capture SN \cite{Nomoto1984}.

Additional evidence for a relatively symmetric supernova comes from the low value measured for the transverse velocity of J0737-3039A/B, $\approx 10$ km s$^{-1}$, which makes it statistically unlikely that SNB provided a large kick to the system \cite{Stairs2006}. This conclusion is additionally supported by the system's position close to the galactic plane \cite{Piran2005}, the low alignment angle between the spin of pulsar A and the orbital angular momentum of the system \cite{Stairs2006,Ferdman2008} and the stability of pulsar A's pulse shape profile \cite{Ferdman2008}.

Under the hypothesis that pulsar B formed from an $e$-capture SN, a constraint on the neutron star equation of state (EoS) can be derived from modeling the progenitor's evolution up to and through the supernova \cite{Podsi2005}. As mentioned above, the progenitor ONeMg core becomes unstable to $e$-capture and collapses at a certain mass. Podsiadlowski et al \cite{Podsi2005} (hereafter referred to as article I) estimated that mass to be in the range 1.366 $M_{\odot} < M < 1.375M_{\odot}$. Under the assumption that negligible mass ($\lesssim 10^{-3} M_{\odot}$) is lost from the core during the SN explosion, this then translates into the range for the baryon mass of pulsar B $M_{\rm B}$. The gravitational mass is measured to be $M_{\rm G}$ = 1.2489$\pm$0.0007 $M_{\odot}$; the difference in the two masses gives the gravitational binding energy in units of mass $BE = M_{\rm B} - M_{\rm G}$. Note that since the gravitational mass is fixed at the measured mass of pulsar B, the binding energy and baryon mass of the neutron star are equivalent quantities. A given EoS predicts a certain binding energy for a neutron star of a given mass; therefore the constraints on $M_{\rm B}$ from modeling and the observed $M_{\rm G}$ together constrain the EoS under the $e$-capture SN hypothesis. Of the EoSs tested in article I, those based on field theoretical descriptions of nucleon-nucleon interactions appeared to give too small a binding energy to be consistent with the constraint. A later study using different field theoretical models found the same systematic under-binding \cite{Blaschke2008}; however it is too early to say whether this is an intrinsic property of EoSs of that type. EoSs with different theoretical underpinnings showed no systematic under- or over-binding with respect to the original constraint, giving instead a scatter of binding energies 1.33$M_{\odot} < M_{\rm B} < 1.38M_{\odot}$. A later simulation of an $e$-capture SN from a $1.377M_{\odot}$ ONeMg core results in a neutron star of baryon mass $1.36 \pm 0.002M_{\odot}$ \cite{Kitaura2006}, indicating core mass loss of up to $10^{-2}M_{\odot}$. Such a mass loss shifts the constraint to lower binding energy, making it consistent with some of the field-theoretical models of the EoS.

A pertinent question to ask is: can we identify specific properties of the neutron star EoS that correlate with the binding energy $BE$ of a neutron star of a given gravitational mass? That is, what EoS property determines whether it will fall within the constraints imposed by modeling the $e$-capture scenario? In this article we shall demonstrate that such a property exists for EoSs of neutron stars that predict a maximum mass above $1.44M_{\odot}$. That property is the first derivative of the nuclear symmetry energy with respect to density at nuclear saturation density $n_0$, commonly denoted $L$. Furthermore, given the correlation between $L$ and $BE$, constraints on $L$ from a variety of terrestrial nuclear experiments allows us to place an independent constraint on the binding energy of pulsar B, and hence ask: is the modeling of the baryon mass of pulsar B under the assumption of $e$-capture SN consistent with nuclear laboratory data?

The article is organised as follows: in section \ref{sec2} we introduce the nuclear symmetry energy and its relation to the neutron star EoS as well as the current constraints on its density dependence from nuclear experiments. In section \ref{sec3} we summarize the wide range of EoSs that we will use to test for a correlation between symmetry energy and baryon mass, and in section \ref{sec4} we demonstrate the correlation between $L$ and the neutron star binding energy, and use this correlation together with the experimental data to constrain the binding energy of pulsar B. In section \ref{sec5} we discuss the implications of this result and in section \ref{sec6} conclude.

\section{\label{sec2}The nuclear symmetry energy}

For completeness, we recall the relation of the nuclear symmetry energy to the neutron star equation of state.

The neutron star equation of state, in the region of neutron, proton, electron, and muon (npe$\mu$) matter is determined by the equation of state of infinite nuclear matter, which may be written as an expansion of the energy per nucleon in terms of the isospin asymmetry parameter $\alpha = 1-2y_{\rm p}$, where $y_{\rm p} = n_{\rm p}/n$ is the proton fraction and $n_{\rm p}$, $n$ the number density of protons and the total baryon number density, respectively:

\be \label{eq1}
E(n,\alpha) = E_{\rm SNM}(n) + \alpha^2 E_{\rm sym}(n) + ...
\ee
\noindent $E_{\rm SNM}(n)$ is the energy per nucleon of symmetric nuclear matter (proton fraction $y_{\rm p}$ = 1/2 or $\alpha$ = 0) and the nuclear symmetry energy $E_{\rm sym}(n)$ is defined

\be \label{eq2}
E_{\rm sym}(n) = \half {\partial^2 E(n,\alpha) \over \partial \alpha^2} \bigg|_{\alpha = 0}.
\ee
\noindent The expansion is in even powers of $\alpha$ only because of the approximate isospin symmetry of the strong force. $\alpha^2 E_{\rm sym}(n)$ is the energy required to convert a certain fraction $\alpha$ of protons into neutrons from symmetric nuclear matter at a given density $n$. The parabolic approximation, in which Eq.(\ref{eq1}) is truncated at second order, is often used, and is generally accurate to high isospin asymmetries ($\alpha \approx 1$) (although some properties of neutron star matter such as the crust-core transition density can be sensitive to higher order corrections \cite{Xu2009}); in that case, from Eq.(\ref{eq1}) the symmetry energy may be written $E_{\rm sym}(n) = E_{\rm PNM} - E_{\rm SNM}$ where $E_{\rm PNM} = E(n,\alpha=1)$ is the energy per nucleon of pure neutron matter.

A fundamental property of symmetric nuclear matter is its saturation at a density of $0.14 \lesssim n_{\rm 0} \lesssim  0.17$fm$^{-3}$ (equivalent to a mass density of $\approx 2.5\times10^{14}$ g cm$^{-3}$), and an energy per particle of $E_0 = E_{\rm SNM} (n_0) \approx -16$MeV. It is customary to expand both the energy of symmetric nuclear matter and the symmetry energy in density around $n_{\rm 0}$. Defining the parameter $x$ to be $x = (n - n_0)/3n_0$, those expansions can be written

\be \label{eq3}
E_{\rm SNM} (n) = E_{\rm SNM} (n_0) + \half K_0 x^2 ,
\ee

\be \label{eq4}
E_{\rm sym} (n) = E_{\rm sym} (n_0) + L x + \half K_{\rm sym} x^2 ,
\ee
\noindent where the parameter $L$ characterizes the slope of the symmetry energy at nuclear saturation density

\be \label{eq6} L = {\partial E_{\rm sym} \over \partial x}\bigg|_{n_0} = 3 n_0 {\partial E_{\rm sym} \over \partial n} \bigg|_{n_0} ,\ee

\noindent and $K_0$ and $K_{\rm sym}$ characterize the curvature of the symmetric nuclear matter energy and the symmetry energy respectively. Note that, because symmetric nuclear matter has a minimum in energy at nuclear saturation density (by definition), there is no first order term in Eq.(\ref{eq3}).

The pressure of infinite nuclear matter is

\be \label{eq8} P(n,\alpha) = n^2 {\partial E(n,\alpha) \over \partial n}. \ee

\noindent Thus, at saturation density, the pressure of nuclear matter can be written

\be \label{eq9} P(n_0,\alpha) = {1 \over 3} \alpha^2 n_0 L, \ee

\noindent within the parabolic approximation. Therefore, at densities around saturation, the dominant contribution to the pressure of nuclear matter comes from the symmetry energy.

Constraints on the magnitude and density dependence of the nuclear symmetry energy have been obtained in analyses of a variety of nuclear experimental data \cite{Li2008,Tsang2009}. Recent modeling of isospin diffusion in heavy ion collisions involving $^{112}$Sn and $^{124}$Sn extract constraints of 62$< L <$107 MeV \cite{Chen2005,Li2005} and 45$< L <$103 MeV \cite{Tsang2009} respectively. The extraction of the range of $L$ is model dependent; the two ranges quoted here, whereas they overlap significantly, are from two different transport model analyses. A study of isoscaling data from multifragmentation reactions yields $L \sim 66$MeV \cite{Shetty2007}. A constraint from the analysis of pygmy dipole resonance (PDR) data gives 27$< L <$60 MeV \cite{Klimkiewicz2007} and from the analysis of the surface symmetry energies of nuclei over a wide range of masses gives 75$< L <$115 MeV \cite{Danielewicz2007}. Finally, the measurement of neutron skins of a wide mass range of nuclei has led to an estimate of 25$< L <$100 MeV \cite{Centelles2009}.

A conservative constraint, therefore, would be the full range given by the above studies 25$< L <$115 MeV, which we shall call $L1$. As an illustrative example, we shall also examine the implications of the constraint of 62$< L <$107 MeV \cite{Chen2005,Li2005} which we shall call $L2$.

\section{\label{sec3}Neutron star equations of state}

A given nuclear matter equation of state completely specifies the EoS of uniform npe$\mu$ neutron star matter (the leptonic contribution to the EoS can be expressed in terms of the symmetry energy through the $\beta$-equilibrium condition \cite{Lattimer2001,Li2008}). In Sec.~\ref{sec2} we defined a number of parameters that characterize the magnitude and density dependence of the energy of nuclear and, hence, neutron star, matter: $E_0, E_{\rm sym} (n_0), L, K_0, K_{\rm sym}$, etc. The sample of EoSs we use in this article span a representative range of these parameters. Our one constraint is that the EoSs used must predict a maximum neutron star gravitational mass of $M_G > 1.44M_{\odot}$ in line with the maximum accurately measured neutron star mass at the current time \cite{Weisberg2005}.

An EoS is constructed using an underlying model of the nucleon-nucleon interaction and a given procedure to calculate the interactions of neutrons and protons in medium. We shall use EoSs with the following theoretical underpinnings (because we are using a large number of EoSs, we give mainly references to works that studied certain subset of them; individual references can be found in those works).

\subsection{Phenomenological}
We use EoSs derived from 63 parametrizations of the nonrelativistic phenomenological Skyrme-Hartree-Fock model. They consist of those classified as group I and group II in \cite{Stone2003} that satisfy our NS mass requirement as well as the BSk6-16 parametrizations \cite{Pearson2009}. Also included are three phenomenological models used in article I: BPAL21, BPAL31 and BAL.

\subsection{RMF}
We use 29 EoSs based on a relativistic mean-field description of the nuclear interaction. We include all 22 used in a previous study \cite{Chen2007} and we direct the reader there for details of those interactions. They include non-linear models, density-dependent meson-nucleon coupling models and point coupling models. We also include the KVR, KVOR \cite{Kolomeitsev2005} and D$^3$C \cite{Danielewicz2002} interactions. Finally, we use four EoSs from article I: GLEN210,GLEN240,GLEN300 and GLENHYB, the first three of which include hyperons and the fourth quark matter in the NS core. The latter four we shall collectively refer to as RMF-Hybrid models.

\subsection{Microscopic}
Six EoSs derived from potential models (based on realistic nucleon-nucleon potentials fit to scattering data together with a phenomenological three-body force) are used. BJ, FPS,APR,WFF and MOSZ \cite{Podsi2005} are calculated using non-relativistic methods whereas the DBHF EoS \cite{Dalen2005} is derived using a relativistic method. BJ, APR, MOSZ and WFF include hyperons at higher densities.

\subsection{MDI}
Also a phenomenological model, the modified Gogny interaction MDI \cite{Das2003} is used. We distinguish this from the other phenomenological EoSs because it was used in the extraction of the constraint on $L$, $L2$, from the isospin diffusion phenomenon \cite{Chen2005,Li2005} and because MDI contains a continuously variable parameter $x$ that directly relates to the slope of the symmetry energy $L$. Each value of $x$ gives a new symmetry energy dependence, and consequently a different EoS, while leaving the symmetric matter EoS the same. This allows us to investigate the effect of varying $L$ while keeping all other EoS parameters fixed at particular values. We generate EoSs by varying $x$ between +0.6 and $-$2.0 ($L = 32$MeV and $L = 153$MeV respectively).

\subsection{QMC}
We use the two EoSs QMC700 and QMC$\pi$4 calculated from the quark-meson coupling model \cite{Stone2007}, a relativistic theory in which nucleons are treated as bags of quarks that couple to meson fields. Hyperons are included self-consistently in these two EoSs.\\

Neutron star models are calculated by solving the general relativistic equations of hydrostatic equilibrium (the TOV equations); pulsar B is rotating slowly enough that corrections because of rotation are negligible. We take the gravitational mass to be fixed at 1.2489$M_{\odot}$. The baryon mass is simply the number of baryons that constitute the star, $A$, multiplied by the mass per baryon $m_{\rm B}$, $M_{\rm B} = A m_{\rm B}$. An important point to note is that to equate that baryon mass with that of the progenitor core in the $e$-capture scenario, the mass per baryon must be taken to be $m_{\rm B} \sim 931.5$MeV/c$^2$, the value appropriate for a ONeMg core, rather than the nucleon mass $\sim 939$ MeV/c$^2$. Taking the latter would give an $\sim$ 0.8\% shift in our baryon mass relative to that obtained by modeling the progenitor collapse, which is a significant difference in our context (e.g., the estimated uncertainty in baryon mass in article I is, 0.011$M_{\odot}$, or 0.8\% relative to the total baryon mass.)

\section{\label{sec4}Dependence of the gravitational binding energy on the slope of the nuclear symmetry energy}

\begin{figure}[!t]
\hspace{1pc}
\centerline{\psfig{file=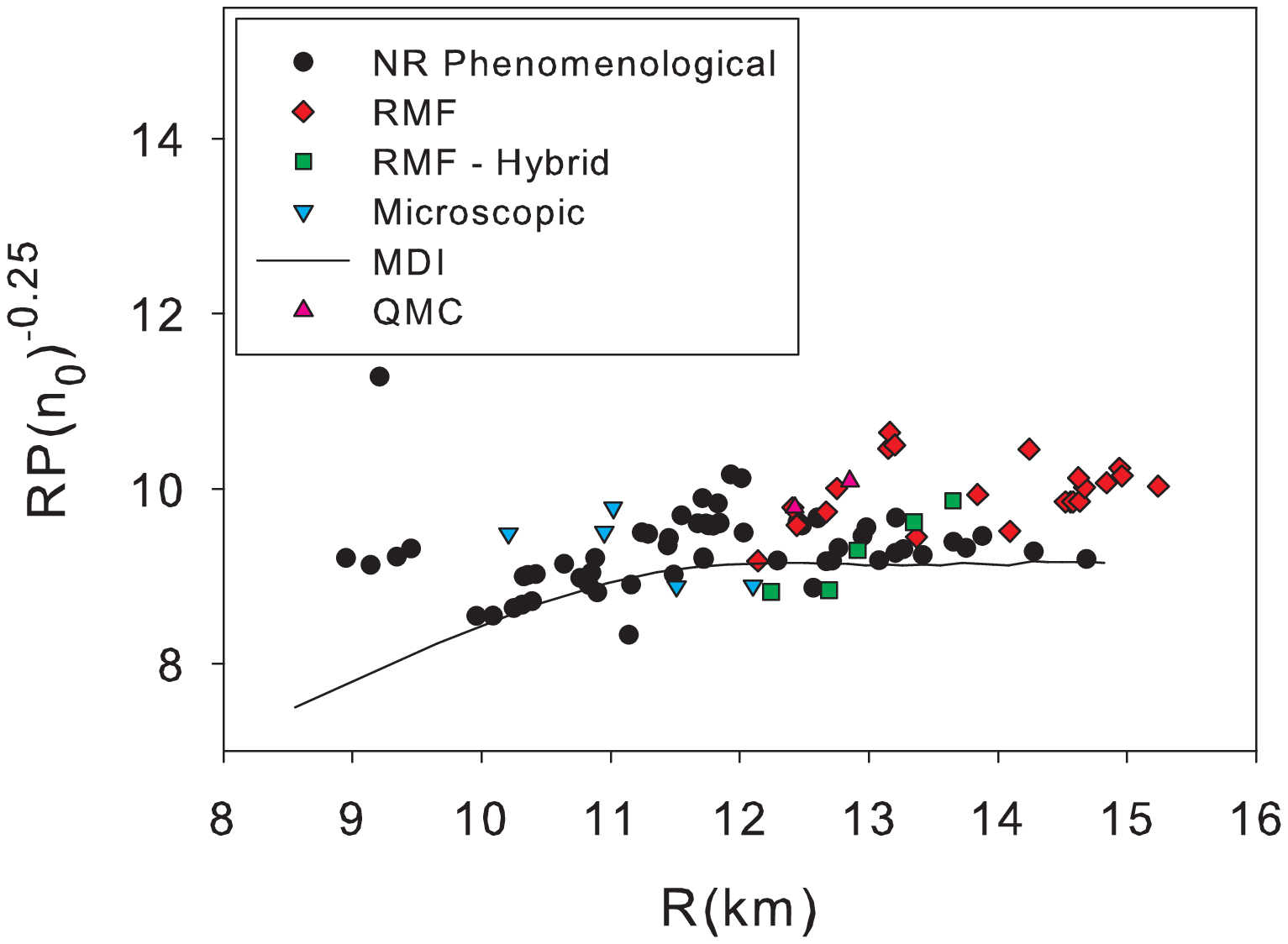,width=10cm}}
\caption{(Color online) The quantity $RP^{-0.25}$ versus neutron star radius $R$ for the sample of EoSs, where the pressure $P$ of neutron star matter is evaluated at saturation density $n_0$.} \hspace{1pc} \label{fig:1}
\end{figure}

It has been demonstrated that there exists a scaling relation between the radius of a neutron star $R$ and its internal pressure $P(n)$ sampled at a baryon number density $n$ = 1-3 $n_0$. Empirically, it was found that for a wide range of neutron star EoSs which predict a maximum neutron star mass $> 1.55 M_{\odot}$ the relation is $RP(n)^{-\alpha} \approx$ const., where $\alpha \approx 0.25$ \cite{Lattimer2001,Steiner2005}. The constant is dependent on the density at which the pressure is sampled, and is weakly dependent on the mass of the star. It was shown that the correlation is stronger the higher the density at which $P$ was evaluated and the lower the mass of the star. We shall evaluate the pressure at saturation density $n_0$. For the EoSs considered in this work, we plot $RP(n_0)^{-0.25}$ versus $R$ in Fig.~\ref{fig:1}; we find $RP(n)^{-0.25} \approx 9.5 \pm 1$. The scatter is somewhat larger than that found in the previous studies \cite{Lattimer2001,Steiner2005}, mainly because we allow EoSs with a lower maximum mass (1.44$M_{\odot}$ rather than 1.55$M_{\odot}$) and evaluate the pressure at a relatively low density (although these factors are compensated to a small extent by the relatively low mass of the neutron star).

\begin{figure}[!t]
\hspace{1pc}
\centerline{\psfig{file=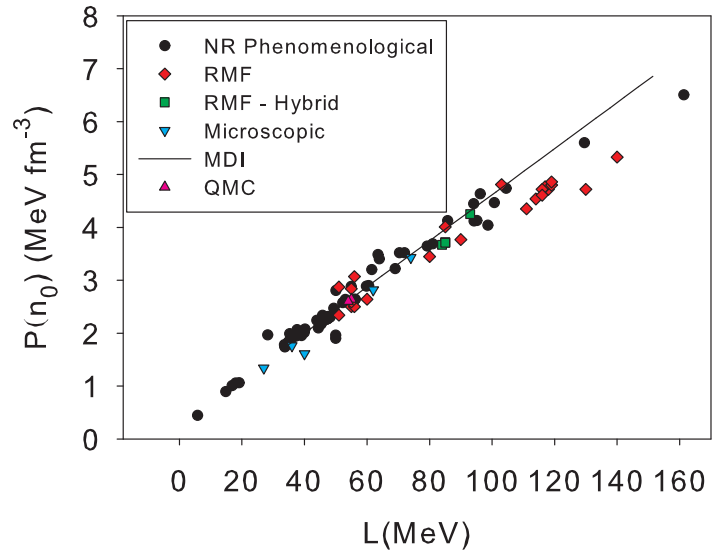,width=10cm}}
\caption{(Color online) The pressure of $\beta$-equilibrium matter at saturation density versus the slope of the symmetry energy $L$ for the sample of EoSs.} \hspace{1pc} \label{fig:2}
\end{figure}

The above correlation can be expressed $P(n_0) \propto R^4$. As discussed in Sec.~\ref{sec2}, the dominant contribution to the pressure of neutron star matter at densities around saturation is from the derivative of the nuclear symmetry energy $L$, as demonstrated in Fig.~\ref{fig:2} for our EoS selection. Indeed, it was pointed out that the $R-P$ correlation translates into an $R-L$ correlation \cite{Steiner2005}: $L \propto R^4$. Let us define the compactness parameter $\beta = G M_{\rm G} / Rc^2$. \emph{At a constant gravitational mass}, we can thus express the correlation $L \propto \beta^{-4}$.

\begin{figure}[!t]
\hspace{1pc}
\centerline{\psfig{file=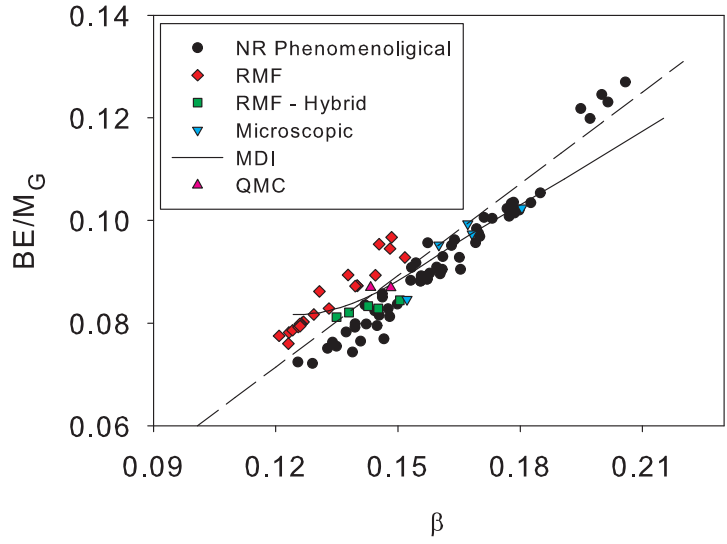,width=10cm}}
\caption{(Color online) The fractional gravitational binding energy of the neutron star $BE/M_{\rm G}$ versus the compactness parameter $\beta$ for the sample of EoSs. The dashed line indicates the analytic Newtonian result.} \hspace{1pc} \label{fig:3}
\end{figure}

The binding energy of a neutron star relative to its gravitational mass, $BE/M_{\rm G}$, is a function of compactness $\beta$ (e.g.,~in Newtonian gravity it is simply $BE/M_{\rm G} =3GM_{\rm G} /  5R = 0.6 \beta$) where $BE$ is in units of mass. In Fig.~\ref{fig:3} we plot $BE/M_{\rm G}$ against $\beta$ for our collection of EoSs. The analytic Newtonian relation is also plotted, and gives a good fit to the binding energy of the relatively low mass neutron star.

To summarize: we have demonstrated for our model neutron star of gravitational mass $M_{\rm G}$ = 1.2489$M_{\odot}$, and for a wide range of EoSs, that the binding energy correlates roughly linearly with the compactness of the neutron star $\beta$ which in turn correlates strongly with the slope of the nuclear symmetry energy $L$ predicted by the EoSs. Therefore we expect to see a correlation between $L$ and the binding energy of the neutron star, and consequently the baryon mass given that the gravitational mass is fixed. In Fig.~\ref{fig:4} we plot $L$ versus the baryon mass of our model neutron star for all EoSs considered; such a correlation is immediately apparent.

\section{\label{sec5}Discussion}

\begin{figure}[!t]
\hspace{1pc}
\centerline{\psfig{file=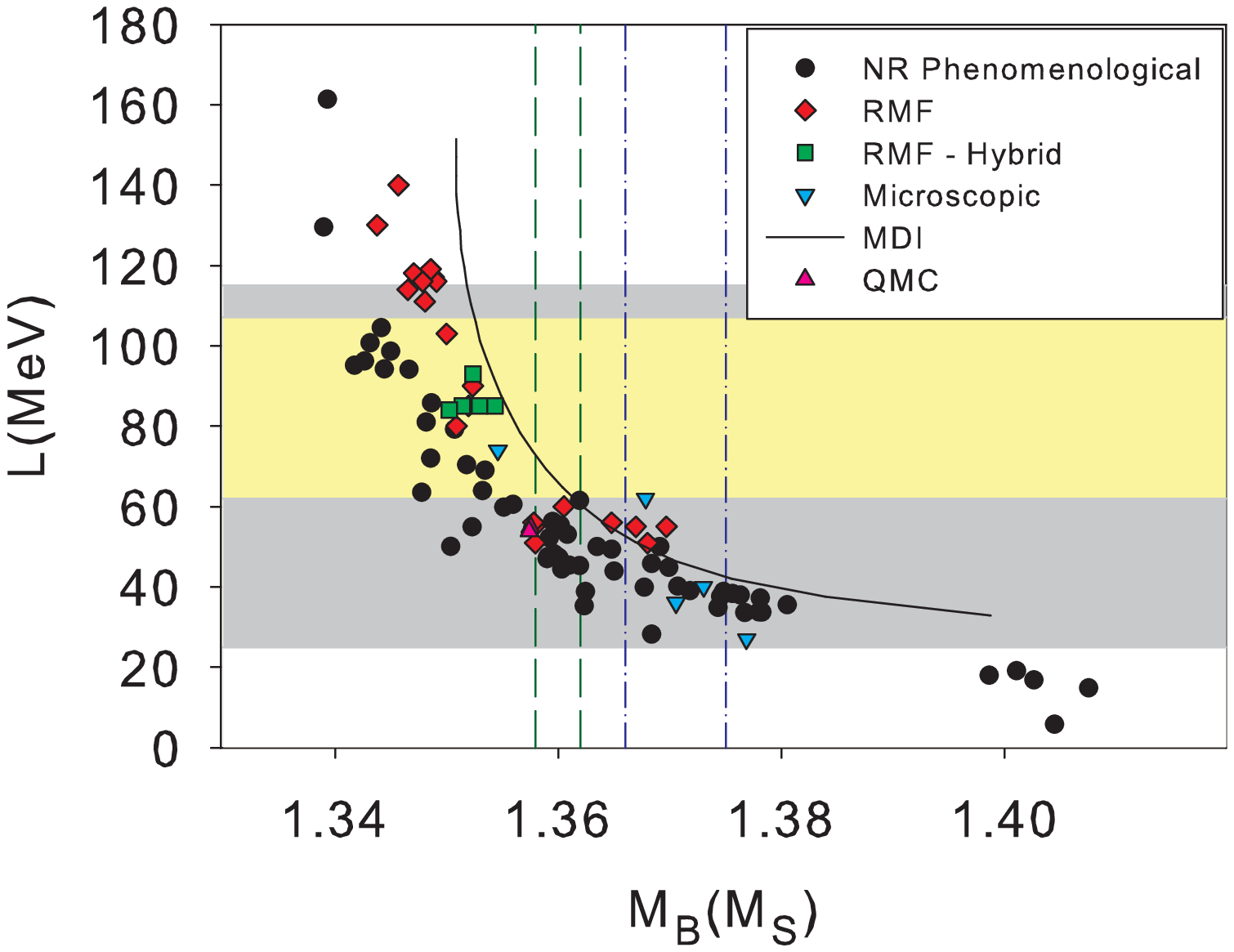,width=10cm}}
\caption{(Color online) The slope of the symmetry energy at saturation density $L$ versus the baryon mass of a neutron star with a gravitational mass of $1.2489M_{\odot}$ for the sample of EoSs. The laboratory constraints on $L$ are represented by the shaded regions; the grey region 25MeV$< L < $115MeV is the most conservative constraint $L1$ whereas the yellow region within is the constraint $L2$. The constraints on the baryon mass $M_{\rm B}$ from modeling pulsar B's progenitor under the assumption of an $e$-capture SN are given by the vertical lines; the dashed lines show the constraint $M1$ and the dash-dot lines the constraint $M2$.} \hspace{1pc} \label{fig:4}
\end{figure}

Fig.~\ref{fig:4} shows that, for a wide range of EoSs with widely varying density dependences and calculated using a number of different theoretical models of the nuclear interaction, a region in $L - M_{\rm B}$ space is preferentially picked out. We shall examine the implications of this correlation in light of the two $e$-capture constraints: $M_{\rm B}$ in the range 1.358-1.362 $M_{\odot}$ \cite{Kitaura2006} and $M_{\rm B}$ in the range 1.366-1.375 $M_{\odot}$ \cite{Podsi2005}, which we shall refer to as $M1$ and $M2$ respectively. In Fig.~\ref{fig:4}, the constraints from nuclear experiment $L1$ and $L2$ are represented by the dark and light shaded horizontal regions, respectively, and the two constraints $M1$ and $M2$ are defined by the dashed and dash-dotted vertical lines, respectively. For the two sets of constraints to show consistency we demand that the region of $L - M_{\rm B}$ space over which they intersect coincide with part of the region populated with EoSs.

Let us first examine how the constraints on $L$, $L1$ and $L2$, constrain $M_{\rm B}$. The most conservative range of $L$, $L1$, is consistent with both $M1$ and $M2$; the constraint $L2$ coming solely from the analysis of isospin diffusion in heavy ion collisions with the MDI interaction is not consistent with $M2$ and only marginally consistent with $M1$. One can see that values of $L$ below $\sim 60$ MeV are consistent with a broad range of $M_{\rm B}$, from $\sim 1.35 M_{\odot}$ up to at least $\sim 1.40 M_{\odot}$; $M_{\rm B}$ is only weakly dependent on $L$ in this range. A value of $L \lesssim 60$ MeV would therefore be consistent with the $e$-capture scenario, but not provide particularly stringent constraints on $M_{\rm B}$ while other properties of the NS EoS remain unconstrained. A value of $L$ above 60 MeV would constrain $M_{\rm B}$ to within $\sim 0.02 M_{\odot}$ at values smaller than the constraint $M2$ and only marginally consistent with $M1$.

One can also consider how the current theoretical constraints from the $e$-capture SN scenario, $M1$ and $M2$, constraint $L$. Taking $M1$ and $M2$ together (giving a conservative range of $M_{\rm B}$ of 1.358 - 1.375$M_{\odot}$), the range of $L$ is constrained to be $L \lesssim$ 70MeV. The $e$-capture scenario thus appears consistent with the \emph{lower} range of $L$ values given by current experimental constraints. Looking at those constraints individually, only the range extracted from the analysis of the surface symmetry energies of nuclei 75$< L <$115 MeV \cite{Danielewicz2007} is completely inconsistent with the $e$-capture scenario.

We finish our discussion with some caveats and remarks. Firstly, we are assuming that pulsar B is a neutron star; that is, at least in the outer parts of the core, it is composed of npe$\mu$ matter. Hyperonic and quark matter may appear in the inner core without altering the results significantly as the binding energy is determined mainly by the physics of matter at saturation density and just above. We also assume that our selection of EoSs span the full range of EoS parameter space consistent with current observation. Regarding the last caveat, note that we are treating all other EoS properties (incompressibility, etc) as unconstrained - we have tested EoSs with as wide a range of those properties as possible. There are however, experimental constraints on some of those properties (e.g., the incompressibility of symmetric nuclear matter at saturation density). Such constraints, and those that come in the future from experiment and observations, may improve the $L - M_{\rm B}$ relation and hence the constraints from $L$. Finally, we have taken the gravitational mass to be 1.2489$M_{\odot}$ ignoring the observational error of $\pm 0.0007M_{\odot}$. This translates into an error of the same magnitude in $M_{\rm B}$, i.e., of $\sim 10^{-3}M_{\odot}$, which does not significantly change any of our conclusions.

\section{\label{sec6}Conclusion}

We have demonstrated that for a fixed gravitational mass $M_{\rm G}$ there exists a correlation between the binding energy, and hence the baryon mass $M_{\rm B}$, of a neutron star and the slope of the nuclear symmetry energy sampled at around nuclear saturation density. The $L - M_{\rm B}$ correlation holds for a wide variety of EoSs calculated within a number of different theoretical frameworks and depends on only two assumptions: (1) that the star is indeed a neutron star and not a pure quark star (e.g. strange star) (2) that the EoS of the neutron star is sufficiently stiff at high densities to give a maximum mass above $1.44M_{\odot}$. Note that we do not exclude the appearance of hyperons, quark matter or other exotica in the denser parts of the star's core provided that they do not soften the EoS such that point (2) is violated.

Combined with all current experimental data on $L$, this correlation gives an approximate constraint on the baryon mass of pulsar B of the double pulsar system PSR J0737-3039 $1.34 \lesssim M_{\rm B} \lesssim 1.4M_{\odot}$. This is an independent constraint on the baryon mass to that derived from modeling the progenitor star to pulsar B under the assumption that it formed in an electron-capture supernova. We can expect the experimental constraints on $L$ to become stricter in the next few years; the properties of isospin-asymmetric systems is at the forefront of experimental investigations at radioactive ion beam facilities (FAIR in Germany, RIKEN in Japan, SPIRAL2/GANIL in France, CSR in China and FRIB in the United States) and the parity-violating electron scattering experiment at JLab (PREX), \cite{Horowitz2001}.

The $e$-capture constraints are shown to be consistent only with a value of $L$ under $\sim$70 MeV, which falls in the lower half of the range constrained from a variety of nuclear experiments. If further experimental data suggests a value of $L$ that is higher, the $e$-capture scenario may have to be revised, for example to include greater mass loss from the progenitor core during supernova, to shift the $M_{\rm B}$ constraint to lower values. The correlation explored in this article reveals an intriguing new mutual interplay between the fields of nuclear experiment, nuclear theory, and astrophysical modeling, which we hope will encourage further experimental investigation and refinement of theoretical models in all areas.

\section*{ACKNOWLEDGEMENTS}
The authors are grateful to Lie-Wen Chen for providing the RMF equations of state, to Jirina Rikovska Stone for useful comments and providing the QMC EoSs, and to Philipp Podsiadlowski and John Miller for useful comments. Finally, we thank the referee for useful comments. This work is supported in part by the U.S. National Science Foundation under Grant Nos. PHY0652548 and PHY0757839, the Research Corporation under Grant No. 7123 and the Texas Coordinating Board of Higher Education under Grant No. 003565-0004-2007.

\newpage
\clearpage

\newpage
\clearpage

\newpage
\clearpage

\newpage
\clearpage

\begin{thebibliography}{99}

\bibitem{Burgay2003}
M. Burgay \emph{et al}, Nature \textbf{426}, 531 (2003)
\bibitem{Lyne2004}
A. Lyne \emph{et al}, Science \textbf{303}, 1153 (2004)
\bibitem{Kramer2008}
M. Kramer and I.H. Stairs, Ann. Rev. Astron. Astrophys. \textbf{46}, 541, (2008)
\bibitem{Dewi2002}
J.D.M. Dewi, O.R.Pols, G.J. Savonije and E.P.J. van den Heuvel, MNRAS \textbf{331}, 1027 (2002)
\bibitem{Ivanova2003}
I. Ivanova, K. Belczynski, V. Kalogera, F.A. Rasio and R.E. Tamm, ApJ \textbf{592}, 475 (2003)
\bibitem{Podsi2005}
Ph. Podsiadlowaki, J.D.M. Dewi, P. Lesaffre, J.C. Miller, W.G. Newton and J.R. Stone, MNRAS \textbf{361}, 1243 (2005)
\bibitem{Nomoto1984}
K. Nomoto, ApJ \textbf{277}, 791 (1984)
\bibitem{Lattimer1989}
J.M. Lattimer and A. Yahil, ApJ \textbf{340}, 426 (1989)
\bibitem{Kitaura2006}
F.S. Kitaura, H.-Th. Janka, W. Hillebrandt, A\&A \textbf{450}, 345 (2006)
\bibitem{Podsi2004}
Ph. Podsiadlowski, N. Langer, A.J.T. Poelarends, S. Rappaport, A. Heger and E. Pfahl, ApJ \textbf{612}, 1044 (2004)
\bibitem{Scheck2004}
L. Scheck, T. Plewa, H.-Th. Janka, K. Kifonidis and E. Muller, PRL \textbf{92}, 011103 (2004)
\bibitem{Stairs2006}
I.H. Stairs, S.E. Thorsett, R.J. Dewey, M. Kramer and C.A. McPhee, MNRAS \textbf{373}, L50 (2006)
\bibitem{Piran2005}
T. Piran and N.J. Shaviv, PRL \textbf{94}, 051102 (2005)
\bibitem{Ferdman2008}
R.D. Ferdman, I.H. Stairs, M. Kramer, R.N. Manchester, A.G. Lyne, R.P. Breton, M.A. McLaughlin, A. Possenti and M. Burgay, AIPC Conf. Proc., \textbf{983}, 474 (2008)
\bibitem{Blaschke2008}
D. Blaschke, T. Klaehn and F. Weber, in Proceedings of the 3rd International Workshop on Astronomy and Relativistic Astrophysics (IWARA), Joao Pessoa, Brazil, 2007, arxiv:0808.1279 (2008)
\bibitem{Xu2009}
Jun Xu, Lie-Wen Chen, Bao-An Li and Hong-Ru Ma, ApJ \textbf{697}, 1549 (2009)
\bibitem{Li2008}
Bao-An Li, Lei-Wen Chen and Che Ming Ko, Phys. Rep. \textbf{464}, 113 (2008)
\bibitem{Tsang2009}
M.B. Tsang, Yingzun Zhang, P. Danielewicz, M. Famiano, Zhuxia Li, W.G. Lynch and A.W. Steiner, PRL \textbf{102}, 122701 (2009)
\bibitem{Chen2005}
Lie-Wen Chen, Che Ming Ko and Bao-An Li, PRL \textbf{94}, 032701 (2005)
\bibitem{Li2005}
Bao-An Li and Lie-Wen Chen, Phys. Rev. \textbf{C72}, 064611 (2005)
\bibitem{Shetty2007}
D.V. Shetty, S.J. Yennello, G.A. Souliotis, Phys. Rev. \textbf{C76}, 024606 (2007)
\bibitem{Klimkiewicz2007}
A. Klimkiewicz \emph{et al}, Phys. Rev. \textbf{C76}, 051603(R) (2007)
\bibitem{Danielewicz2007}
P. Danielewicz and J. Lee, AIPC Conf. Proc. \textbf{947}, 301 (2007)
\bibitem{Centelles2009}
M. Centelles, X. Roca-Maza, X. Vinas and M. Warda, PRL \textbf{102}, 122502 (2009)
\bibitem{Lattimer2001}
J.M. Lattimer and M. Prakash, ApJ \textbf{550}, 426 (2001)
\bibitem{Weisberg2005}
J.M. Weisberg and J.H. Taylor, Astron. Soc. Pac. Conf. Ser. \textbf{328}, 25 (2005)
\bibitem{Stone2003}
J. R. Stone, J.C. Miller, R. Koncewicz, P.D. Stevenson and M.R. Strayer, Phys. Rev. \textbf{C68}, 034324 (2003)
\bibitem{Pearson2009}
J.M. Pearson, S. Goriely, N. Chamel, M. Samyn and M. Onsi, AIPC Conf. Proc. \textbf{1128}, 29 (2009)
\bibitem{Chen2007}
Lie-Wen Chen, Che Ming Ko and Bao-An Li, Phys. Rev. \textbf{C76}, 054316 (2007)
\bibitem{Kolomeitsev2005}
E.E. Kolomeitsev and D.N. Voskresensky, Nucl. Phys. \textbf{A759}, 373 (2005)
\bibitem{Danielewicz2002}
P. Danielewicz, R. Lacey and W.G. Lynch, Science \textbf{298}, 1592 (2002)
\bibitem{Dalen2005}
E.N.E. van Dalen, C. Fuchs and A. Faessler, Phys. Rev. \textbf{C72}, 065803 (2005)
\bibitem{Das2003}
C.B. Das, S. Das Gupta, C. Gale and Bao-An Li, Phys. Rev. \textbf{67}, 034611 (2003)
\bibitem{Stone2007}
J. Rikovska Stone, P.A.M. Guichon, H.H. Matevosyan and A.W. Thomas, Nucl. Phys. \textbf{A792}, 341 (2007)
\bibitem{Steiner2005}
A.W. Steiner, M. Prakash, J.M. Lattimer and P.J. Ellis, Phys. Rep. \textbf{411} 325 (2005)
\bibitem{Horowitz2001}
C.J. Horowitz, S.J. Pollock, P.A. Souder and R. Michaels, Phys. Rev. \textbf{C63} 025501 (2001)

\end{thebibliography}
\end{document}